Optimizing Complex Health Intervention Packages through the Learn-As-you-GO (LAGO) Design

Donna Spiegelman, Center on Methods for Implementation and Prevention Science and Department of Biostatistics, Yale School of Public Health, New Haven, CT, USA

Dong (Roman) Xu, Southern Medical University Institute for Global Health (SIGHT), Dermatology Hospital, Southern Medical University, Guangzhou, China; Acacia Lab for Implementation Science, Department of Health Management, School of Health Management of Southern Medical University, Guangzhou, China

Ante Bing, Department of Mathematics and Statistics, Boston University, Boston, MA, USA

Guangyu Tong, Section of Cardiovascular Medicine, Department of Internal Medicine, Yale School of Medicine, New Haven, CT, USA

Mona Abdo, Center on Methods for Implementation and Prevention Science and Department of Biostatistics, Yale School of Public Health, New Haven, CT, USA

Jingyu Cui, Center on Methods for Implementation and Prevention Science and Department of Biostatistics, Yale School of Public Health, New Haven, CT, USA

Charles Goss, Center for Biostatistics and Data Science, Washington University School of Medicine, Saint Louis, MO, USA

John Baptist Kiggundu, Infectious Diseases Research Collaboration, Kampala, Uganda

Chris T. Longenecker, Division of Cardiology & Department of Global Health, University of Washington, Seattle, Washington, USA

LaRon Nelson, Yale School of Nursing, New Haven, CT, USA

Drew Cameron, Department of Health Policy and Management, Yale School of Public Health, New Haven, CT, USA

Fred Semitala, Infectious Diseases Research Collaboration, and Department of Medicine, Makerere University, and Makerere University Joint AIDS Program, Kampala, Uganda

Xin Zhou, Center on Methods for Implementation and Prevention Science and Department of Biostatistics, Yale School of Public Health, New Haven, CT, USA

Judith J. Lok, Department of Mathematics and Statistics, Boston University, Boston, MA, USA

Abstract

In the face of vast numbers of preventable deaths worldwide and gaping disparities in their distribution, we cannot afford to conduct null and inconclusive effectiveness and implementation trials of evidence-based interventions. The gold standard in biomedical research, the individually randomized clinical trial, is ill-suited as the primary tool for knowledge generation for contextually relevant, scalable, complex public health interventions of multi-component strategies. In this paper, we discuss the new Learn-As-you-GO (LAGO) design. In LAGO trials, the components of a complex intervention package are repeatedly optimized in pre-planned stages, until the package achieves its outcome and power goals with minimized cost and/or other optimization criteria, such as maximizing patient satisfaction. In this paper, the inputs to, and outputs of, LAGO are described, along with its general methodology. The methods are illustrated in the BetterBirth study, a large trial that aimed to reduce maternal and neonatal mortality in Uttar Pradesh, India, using WHO's essential birth practices checklist. Despite its scale, the BetterBirth study failed to demonstrate a significant effect of the intervention package on the primary health endpoint that included maternal mortality.  We show how this unfortunate outcome could have been remedied had LAGO been used. LAGO is further illustrated through the discussion of several ongoing LAGO-informed implementation trials of HIV and non-communicable diseases in the United States and Sub-Saharan Africa.

The Learn-As-you-GO (LAGO) design optimizes a complex, multi-level intervention for minimum cost, pre-specified power, and a pre-specified effectiveness goal, by adapting the intervention as the study is conducted, reducing risk of trial failure.

In 2014, Atul Gawande, a physician, MacArthur Genius Award winner, and Assistant Administrator for Global Health at USAID between 1922-1925, turned his attention towards reducing maternal and neonatal mortality.   Neonatal mortality is one of the world's greatest global health inequities and the focus of one of the World Health Organization's 8 Millenium Development Goals which remains a focus of its active Sustainable Development Goals. Gawande advocated for the adoption and implementation of a similar Safe Childbirth Checklist (SCC)[2]. This checklist is based on 28 WHO-recommended essential birth practices (EBP). It was evaluated in a massive cluster-randomized trial involving almost 160,000 mothers in Uttar Pradesh, India, a region that has had alarmingly high neonatal and maternal mortality rates. The BetterBirth trial implemented a multifaceted implementation strategy, the SCC, that operationalized a bundle of strategies to facilitate its use. The bundle included (1) a 2-3 day educational and motivational workshop (the launch), (2) 43 days of peer coaching visits for birth attendants at the intervention sites, (3) identification of coach team leaders, who accompanied peer coaches on half of their visits, and (4) appointment of a local champion for use of the checklist and as the Childbirth Quality Coordinator. Five years later, the BetterBirth results were reported in the *New England Journal of Medicine*, showing no significant effect on the primary health outcome[3-5].

How could this have happened? About this, much has been written, with at least 299 citations as of this writing[6-9]. Although the study was not formally defined as a hybrid design, it bears similarity to a Hybrid Type 1 design[10], as both the health intervention and its implementation outcomes were evaluated, with the clinical outcomes considered primary. As such, the trial's final report highlighted that the implementation of the EBP strategies failed to achieve sufficiently high adoption and fidelity to measurably improve health outcomes[11]. An examination of the experience of the BetterBirth trial underscores the tension between the rigid structure of randomized clinical trials and the dynamic nature of real-world implementations. The BetterBirth trial's implementation strategy bundle consisted of a complex set of components aimed at promoting EBPs—the evidence-based intervention (EBI) of this study—and were fixed at baseline, with no contextual adaptions allowed. This static approach is central to the traditional methodology of randomized clinical trials. Dr. V, a member of the BetterBirth Data Safety and Monitoring Board (DSMB), shared his experience: "It was heartbreaking to sit on that DSMB, watching the study fail, and have our hands tied – there was nothing we could do." Had the investigators been able to adapt, tweak, or tailor the intervention as the trial progressed, they likely could have revised the implementation package to address the barriers and challenges that were observed as the trial progressed to prevent the successful implementation of this multi-component complex bundle from adequately impacting the targeted health outcomes.

The 'voltage drop', that is, the decline in estimated impact of an intervention as the evaluation progresses from efficacy to effectiveness to implementation, is a well-established feature of implementation research [12]. The failure rates for implementing complex innovations in healthcare organizations have been reported to range between 30% to 90% depending on the scope of the organizational change involved, the definition of failure, and the criteria to judge it[13]. Similar high rates of failure with respect to targeted clinical outcomes have been reported for quality improvement projects[14]. The failed SPRINT MIND trial[15] exemplifies this phenomenon in the pragmatic trial domain. This trial, that aimed to reduce dementia through blood pressure control, was stopped early due to overwhelming evidence of benefit for the combined cardiovascular disease endpoint of the trial at the expense of ending further learning about the impact of blood pressure control on dementia.  The Learn-As-you-GO (LAGO) design allows investigators to adapt their complex, multi-component intervention package to optimize its effectiveness and achieve a particular outcome goal, preserve statistical power, and/or minimize cost[1,16,17].  With Learn-As-you-GO (LAGO), had there not been a competing endpoint taking higher priority, it might have been possible to increase the treatment arm dose in planned stages to optimize efficacy against safety, thereby 'saving' the trial and further producing knowledge for improved clinical care for dementia. In another example, the ADAPTABLE (Aspirin Dosing: A Patient-centric Trial Assessing Benefits and Long-term Effectiveness) trial was a three-year pragmatic trial comparing the effectiveness of two different daily doses of aspirin widely used to prevent heart attacks and strokes in individuals living with heart disease[18]. No significant differences in the health outcomes or side effects were found between the two groups.  With LAGO, the trial would not be limited to study only the two doses chosen but could optimize the dose across a much wider range of possibilities for a specific cardiovascular disease prevention goal, subject to, say, minimizing side effects, perhaps further optimized by patient sub-groups, identifying a lower effective dose with fewer side effects.

Standard methods for designing and conducting randomized clinical trials prohibit adaptation, tailoring or tweaking of the intervention strategy bundle in ongoing trials, even in the face of unfolding programmatic failure. The culture, values, and methodologies of randomized clinical trials, which have proven to be extraordinarily useful for knowledge generation in drug and device development, may often not be optimal when applied rigidly to implementation trials of complex multi-level multi-component implementation strategy bundles, whether randomized, often by large clusters as in BetterBirth, or not. Although adaptation, tailoring and tweaking are acknowledged as phenomena that invariably occur and a framework has been proposed to record these changes, until now, no methods have been available to account for these changes in the subsequent statistical analysis[12,19,20]. For such trials, a Learn-As-you-GO (LAGO) design could be more suitable, as it accommodates ongoing learning and adaptation in a quantitatively rigorous manner while maintaining the underlying statistical integrity of the study[1,16]. This article provides an overview of the LAGO approach, outlining the steps for conducting a LAGO study and providing several illustrative examples, including a hypothetical application to the BetterBirth trial.

**Development of LAGO**

The need for a Learn-As-you-GO (LAGO) design that allows researchers to optimize complex, multicomponent interventions was first pointed out to this paper's first author, Dr. Donna Spiegelman, by physician and implementation scientist, Dr. Lisa Hirshhorn, now at Northwestern University, one of the original BetterBirth investigators. It took quite some time and the engagement of an expert mathematical statistician, Dr. Judith Lok (Boston University) , along with Spiegelman's post-doctoral research fellow at the time, Dr. Daniel Nevo  (Tel Aviv University), to develop a solution that solved this problem in a mathematically rigorous manner, alongside at least one extension to accommodate a greater range of data types[1,16].

LAGO can be applied to hybrid designs I, II, and III in implementation research, as well as to pragmatic trials and trials of new investigational drugs and devices with a multi-component feature, such as multi-drug regimens, e.g. anti-retroviral therapy (ART) for HIV/AIDS treatment[10]. In implementation science, these designs are used to evaluate an EBI or a package of EBIs, as well as their corresponding implementation strategies, all of which may be multicomponent or, at least, have a range of possible doses. LAGO can be used to adapt and optimize the components of the EBI or implementation strategies, or both, depending on research needs. LAGO trials have multiple stages. First, researchers develop the initial intervention package. After each subsequent stage, new doses of the intervention components are recommended aimed at achieving pre-defined optimization criteria, such as minimizing cost while attaining a pre-specified effectiveness target, or maximizing patient satisfaction within a pre-specified budget. This process continues through subsequent stages, described in detail below. The final analysis uses data from all stages combined.

The LAGO approach follows these steps (Figure 1):

1. **Develop a theory of change and determine the initial intervention package[21]**: In LAGO, based on subject matter knowledge, pilot studies when available, and the literature, researchers propose or co-produce with stakeholders[22] an initial multi-component intervention package, mapping out the hypothesized relationships between the implementation components, mediators, and outcomes, potentially through the development of a logic model[23]. The intervention package may include one or more EBIs to be implemented and/or strategies to facilitate their implementation. In standard implementation research, the initial intervention package is fixed at baseline and, formally, would not change, regardless of what is observed as the study progresses.

2. **Determine optimization criteria**: In implementation science, optimization does not seek an absolute "best" but rather the most advantageous under pre-specified, contextually relevant constraints. Thus, the LAGO process begins by establishing clear optimization criteria, which should include specific objectives (outcomes) to be achieved within the constraints. Optimization objectives may involve a minimally acceptable change in health outcomes, similar to the traditional minimum clinically important difference, cost-effectiveness, and/or patient satisfaction. These objectives must be met while adhering to key constraints, typically related to resource inputs such as a predetermined budget or human resource availability. In addition, minimally and maximally feasible bounds on each component must be given.

3. **Costing**: Although not required for LAGO, whenever the optimization objectives include metrics related to cost-effectiveness, which has been the case thus far in our applications of LAGO, with the assistance of health economists, the total implementation cost as a function of implementation intensity for each component are estimated. For some components, values can be obtained from the literature and adjusted for the cost of living between the data sources and the current study sites. Often, costing is already a part of the implementation trial since cost-effectiveness is an essential requirement for sustainability[24].

4. **The initial stage of implementation**: In the start of Stage 1, participants are enrolled to receive the initial intervention package developed before the trial starts, typically by relatively ad hoc methods. Implementation is typically cluster-based (for example, at a health facility, school, etc.). Site-specific tailoring may occur and should be recorded as it will inform the subsequent optimizations[20]. Data are collected on the actual implementation of the intervention components, the outcomes targeted for optimization, and other related contextual factors, particularly those factors for which customized optimization may be of interest.
5. **Iterating through subsequent LAGO stages**: At the end of Stage 1, a model including all intervention components is fitted to predict the optimization outcome(s). Coupled with pre-specified optimization criteria, the estimated coefficients of the outcome model are used in the LAGO optimization to recommend intervention components and their levels to be implemented in Stage 2. This process continues across subsequent stages, with the outcome model fit based on the data from all previous stages combined. The number of stages can vary as study resources permit, or can conclude when the recommended optimized package has stabilized. In our experience thus far, 2-3 LAGO stages have been employed and denote the limits of what is practically feasible. Using the qualitative data collected to date, the recommended intervention at each stage can be adapted to incorporate this additional, rigorously analyzed information to understand the determinants and processes of implementation, and, potentially, to modify the subsequent quantitative recommended implementation strategy bundle in Stage 2 and beyond. The LAGO optimizations should be planned and specified pre-trial.

   At the end of each stage, the independent effects of each component in relation to the optimization outcome(s) are re-estimated using combined data from all stages completed to date and used to calculate the expected values of the optimization goals and constraints to inform the next stage optimization. (e.g., data from Stages 1 and 2 are combined to produce Stage 3 recommendations)
6. **Final analysis**: The basic intuitive idea is that *under the null hypothesis* of no effect of any of the intervention components, within each cluster, the LAGO optimizations do not affect the distribution of the outcomes. All aspects of the final analysis use the data from all stages combined. The final estimate of the optimal intervention package is calculated, along with a 95% confidence set of the possible intervention packages that satisfy the optimization goals within the constraints to express the uncertainty in the optimal package. The optimal package can be tailored to subgroups, e.g., rural vs. urban, large vs. small facility. As in any trial, the standard test of the overall complex intervention framework is conducted, which maintains its usual Type I error rate, as proven in the previous publications[1,16]. Additionally, researchers will obtain estimates and 95% confidence intervals for the effects of each individual intervention component, and estimates and 95% confidence intervals for the expected value of the effectiveness goal(s) of the optimal estimated intervention package.

Boxes 1 and 2 concisely summarize the required inputs and all outputs of a LAGO study, exemplified by the BetterBirth study described next.

**Worked Illustrative Example – The BetterBirth study**

Let's make LAGO more concrete, and consider what could have been done during the BetterBirth trial had it followed a LAGO design. The following analysis is adapted from one of our previous publications[1]. As discussed above, the LAGO design integrates data from multiple stages to repeatedly refine and optimize the intervention package. In this hypothetical example, we aimed to have at least 80% of the EBPS measured at each stage to have been implemented. There was no clinical effectiveness outcome in this illustrative optimization. Outcome data on the percentage of EBPs performed was collected before and after implementation in 3 stages, from 2 facilities in Stage 1, 4 different facilities in Stage 2, and in Stage 3, 15 facilities each in the control and intervention arms, all different facilities from the previous ones. The control arm and baseline periods followed standard of care. The observed percentages of EBPs performed out of those measured by the end of each stage were 38%, 36%, and 69% in the intervention periods, compared to 21%, 17%, and 40%, respectively, in the control periods, indicating substantial improvements but still not attaining the goal of 80%. Nevertheless, the overall effect of the implementation strategy bundle on adherence to EBPs was highly significant (p<0.001).

In this illustrative example, we optimized a two-component implementation strategy bundle consisting of (1) Launch Duration - the number of days dedicated to initiating the intervention at a facility; and (2) Coaching Visits - the cumulative number of follow-up visits needed at a facility to reinforce the practices. To estimate the optimal intervention package, we exploited the naturally occurring variation in the launch duration and number of coaching visits across the study's three stages of development. In Stage 1 of BetterBirth, the launch lasted 3 days, and in Stages 2-3, it was 2 days. The actual mean number of coaching visits across the facilities increased from an intended 1 visit/week (0.72/week actual) to a planned 3 visits/week (1.6 actual) to a planned 2 visits/week (0.92 actual), more frequently at first and tapering off to once monthly.

We re-analyzed BetterBirth to ascertain how different the design might have been had LAGO been used to choose the implementation strategy doses at Stages 2 and 3, using the actual study data at the end of each stage (Table 1). To conduct these two optimizations, the data up to the current stage were used to estimate the impact of each implementation strategy on the proportion of EBPs performed out of those measured, adjusting for birth volume of the facility (births/month) as an example of an a priori facility characteristic that might necessitate tailored strategies. As seen in Table 2 (adapted from[25]), the estimates did not stabilize to barely changing values until Stage 3. Both strategies had a significant impact on the proportion of EBPs performed out of all possible birth practices measured (p < 0.001), with each launch day increasing the EBP uptake by 18 percentage points, and every 5 of coaching visits by 19 percentage points. Using this information, we sought to identify the cheapest intervention package that could be expected to achieve an implementation outcome goal of 80% EBPs performed. The total implementation cost as a function of implementation intensity were taken to be $800 per launch day ($x_1$) and $170 per coaching visit ($x_2$). Cubic cost functions were considered for each strategy (Figure 2), allowing for an economy of scale at lower doses of intervention components, where increasing the number of launch days or coaching visits has less impact on the total cost, and building in prohibitive costs as the upper limits of the intervention component bounds are approached.

Assuming that the launch duration can range between 1 and 5 days while between 1 and 40 coaching visits are a feasible range for that strategy, the recommended cost-effective implementation strategy bundle for achieving a mean of performed EBPs of at least 80% for a facility with an average birth volume of 175 births per month was a launch duration of 4 days and 36 coaching visits. The estimated optimal intervention package was found through considering all possible bundles within the feasibility bounds, and choosing the cheapest one among all of those that were estimated to lead to at least 80% of EBPs performed. For bundles with larger numbers of strategies to be optimized, we have found that numerical optimization methods are more computationally efficient. Under this estimated optimal implementation strategy bundle, the 95% confidence interval for the proportion of EBPs performed was 78% - 82%.

Our analysis also produced a 95% confidence set of alternative combinations of launch duration and coaching visits that might also have achieved EBPs performed that fell within the 95% confidence interval of the predicted outcome goal given by the optimal intervention. Alternative implementation strategy doses in this confidence set were $(3, \geq 39)$, $(4, 33 - 37)$ and $(5, 27 - 34)$, where the first number is the number of launch days and the second the number of days of coaching visits. This confidence set contains 7.5% of all possible implementation strategy bundles contained within the bounds. The estimated cost of the estimated optimal intervention package of 4 launch days and 36 coaching visit days was $16,250/facility (interquartile range $16,324-$18,545, as given by the minimum and maximum cost of the 95% confidence set of the implementation strategy bundles).

Uncertainty in each of these LAGO outputs arises from the uncertainty in estimating the effects of the individual implementation strategies in the outcome model. Once the LAGO study is complete, the estimated coefficients for implementation strategies can be used, along with other inputs, to recommend optimal intervention packages in new settings. The total implementation cost function in relation implementation strategy intensity can also be revised, including its form. The only requirement for the determination of optimal strategy bundle doses in new settings is that the covariates defining the subgroups need to have been included in the outcome model, along with any interactions between these covariates and intervention components as needed. We recommend strict criteria for retention of interaction terms in these models, as these are rarely transportable to new settings [26,27]. Inputs given in Box 1 below can also be modified to recommend optimal interventions in new settings, as illustrated in the next section.

**Towards the development of "personalized" public health interventions through LAGO**

In the example above, to illustrate 'personalization' of the optimal implementation strategy bundle, we estimated it for facilities with 175 births per month, the average monthly birth volume across the BetterBirth facilities. The optimal intervention package could have been estimated for facilities with any other birth volume; for example, the optimal bundle for a low birth volume facility, set at the 25[th] percentile, 130 births per month, was estimated as a 4 day launch duration and 33 coaching visits, not very different from the optimal package at the mean birth volume, suggesting in this case low sensitivity to facility size. Because the optimal intervention package rule is one of the outputs of the LAGO design, a customized optimal intervention package can be given for facilities based upon their measured numerical characteristics used to estimate the rule, whether or not the actual values of the characteristics measured were observed in the study. The recommended Estimated intervention package from a previous LAGO study can thus be used as the starting value for a new LAGO study, or even as the intervention package fixed at baseline in a new study, in a new setting or at a larger scale. Implementation at a larger scale may involve extrapolating outside the range of the observed data and can be risky, but in real-life public health settings it might nevertheless be best to use prior empirical data, in addition to common sense, expert advice, and contextual considerations, to guide scale-up and scale-out.

**Variations of LAGO designs**

There are several versions of the LAGO design. LAGO studies may include both a control group and/or control periods, as well as planned or unplanned variation. In a *controlled* LAGO design, there is an initial randomization of clusters or individuals to the group that does not receive the intervention, which typically continues to receive standard of care, and to the intervention group, with both groups typically consisting of half the study clusters. In BetterBirth, stage 3 had this design. Alternatively, or in addition, there may be concurrent variation in the intervention package component doses. This variation may be planned, and, ideally, randomized, or unplanned. In our experience thus far, if this variation in the intervention component doses is unplanned, this LAGO design takes on a quasi-experimental flavor, where within-cluster comparisons will generally provide full control for time-invariant confounding as is typically the case with repeated measures data[11]. For time-varying confounding, causal inference methods can be used in the stage-wise and final analyses to allow for rigorous inference and subsequent LAGO optimization. It is our experience with implementation studies currently practicing LAGO that there is substantial variation in the doses of intervention components around the recommended values, within clusters across time, as well as between clusters at any given time.

**Examples of ongoing studies employing LAGO designs**

The first author of this paper introduced a consideration of LAGO to the NIMH-sponsored Enabling translation of Science to Service to ENhance Depression CarE (ESSENCE) trial led by Vikram Patel of Harvard and Sangath – India [3,28]. To fill the overwhelming need for mental health professionals in India, a country of 1.5 billion people with less than 10,000 mental health professionals, it has been shown that, after a short face-to-face training, non-specialized health workers can successfully conduct depression screening and treatment through a brief psychological intervention, the Healthy Activity Program. Unfortunately, even with this level of task shifting, it is prohibitively expensive in terms of time and money to offer face-to-face training to enough of India's community health workers to meet the overwhelming need for mental health services. Thus, a non-inferiority trial has been rolled out to compare the effectiveness and cost-effectiveness of face-to-face training to a digital alternative. The digital alternative can be viewed as a multicomponent implementation strategy bundle, since, to maximize effectiveness, components of the curriculum can be tweaked in many ways: reading level, amount of video content, frequency and number of quizzes, digital content and graphics to promote engagement, intensity of the training period, and duration of the training period.

If MOST were to be used in ESSENCE, in Phase 2 of the Multiphase Optimization STrategy (MOST) framework, a factorial design would be used to identify components of the training program with effective doses[4]. With so many components to potentially optimize, a fractional factorial design can be used to ensure that there is sufficient unconfounded variation to estimate the effects of each individual component. In MOST Phase 3, a full trial is conducted to test the effectiveness of the bundle of components found in Phase 2 to be the most effective, compared to no training, or, alternatively, the effectiveness of two alternative bundles can be compared against each other. At baseline, combinations of the alternative doses of components can form unique implementation strategy bundles, which are randomized to trainees or groups of trainees. At the end of the study, the independent effect of each component can be estimated at its fixed baseline dose, given sufficient planned variation in the components assigned. For example, at the end of such a study, we would be able to assess the evidence for the effectiveness of training materials produced at a 5th grade reading level, compared to no training materials at all, but we would not know the optimal reading level. Similarly, at the end of the study, we would be able to evaluate whether a training program in which 25% of the training time was spent watching training videos was effective, compared to no training videos at all. Alternatively, a LAGO design could be used in an enhanced MOST Phase 3 to further optimize the initial implementation strategy bundle selected by MOST Phase 2 and/or by investigators' expert judgements. This approach allows successive LAGO stages to further optimize the component doses while the full randomized trial is underway, typically taking advantage of unplanned variation in doses. In addition, if the initial strategy bundle turns out to be too weak to produce the hypothesized overall intervention effect, when optimizing for power as well as impact, LAGO will reduce the chance that this trial will fail, in the presence of the possible 'voltage drop' discussed in the Introduction[12,29].

LAGO is incorporated into several ongoing clinical trials, including PULESA-Uganda, a 30-month trial running between 2023 and 2026[30]. This stepped-wedge cluster randomized trial in the Kampala and Wakiso districts of Uganda is comparing the effectiveness of two bundles of implementation strategies for the delivery of integrated HIV-Hypertension care into routine practice, a basic and an enhanced version (HTN-BASIC and HTN-PLUS). PULESA was powered to achieve a 34% hypertension control rate by the end of the study, compared to a baseline control rate of 22%, equivalent to an intervention odds ratio of 1.83, a modest goal that is expected to be obtainable and adequately powered given the large size of this study. The HTN-PLUS implementation strategy bundle includes 6 components. Before the start of the first LAGO stage, it is advisable, although not required, to ``guesstimate'' the value of the odds ratio for each package component, with higher values reflecting stronger effects. Then, by multiplying the expected odds ratios of the individual implementation strategies by the baseline doses for each, the feasibility of the initial intervention package to achieve the study's clinical outcome goal can be assessed. Here, at the hypothesized effects of the 6 components at their initial doses as given in Table 3, the overall intervention effect would correspond to an odds ratio of 5.31, leading to a projected 60% of patients under hypertension control, should baseline control rates be 22% as projected before the start of the trial. Note that an odds ratio of 5 is typically quite an ambitious goal for a study intervention.

These sorts of calculations can indicate that an implementation study may have unrealistic study goals, leading to further design modifications before starting the study.

To construct cost-effective implementation strategy bundles using LAGO, the total implementation cost function in relation to program implementation intensity must be known or estimated. Cost-effectiveness is often considered essential for ensuring sustainability[31]. In PULESA, optimizations occurred at 9 and 18 months following the start of follow-up. Data collected up to each time point was used to optimize the presence and dose of pre-specified components of the HTN-PLUS package. At the start of stages 2 and 3, a recommended estimated optimal cost-effective intervention package was suggested. Optimization was designed to minimize the cost of the overall package, giving 80% power to test the null hypothesis of no effect of HTN-PLUS vs. the control condition and given that the pre-specified goal of 34% population-level hypertension control (vs. 22%) would be achieved. It is important to note that LAGO optimizations are not interim effectiveness analyses, and we have shown that there is no need to adjust the overall Type I error rate for the primary effectiveness analysis[16,32]. Unlike an interim analysis in the traditional sense, no testing of the intervention effect is conducted, and the trial data are used only to recommend the optimal intervention package for the next stage[7,16,25].

Similar to what we describe for PULESA, we are also applying LAGO to two other trials participating in the NHLBI's Heart, Lung, and Blood Co-morbiditieS IMplementation Models in People Living with HIV (HLB-SIMPLe) Alliance. These projects focus on integration of hypertension care into HIV clinics in 6 sub-Saharan African countries, TASKPEN-Zambia[33] and MAP-IT Nigeria[34].

LAGO is also planned to be used in a two-year HIV prevention trial, HPTN 096: Getting to Zero among Black Men Who Have Sex with Men (MSM) in the American South[35]. HPTN 096 will use an interrupted time series design to evaluate the efficacy of a multilevel multi-component implementation strategy that includes a combination of community-, organizational-, and interpersonal-level components, designed to increase rates of HIV pre-exposure prophylaxis (PrEP) uptake among Black MSM living without HIV, and medical care retention rates among Black MSM living with HIV. Cross-sectional baseline and endpoint assessments will be conducted at the start and the end of the 2-year intervention period. Proposed LAGO components are presented in Table 4, with definition and measures, units, and lower and upper bounds of components defined. The odds ratios of each component in increasing the PrEP prescribing rates in implementing healthcare facilities are guesstimated. Nine months into the study, the LAGO optimization will leverage the interim PrEP prescribing outcome data measured through electronic health records at participating healthcare facilities. Here, 10 implementation strategies are being considered for optimization: education (proportion of participants who attend at least 1 coalition activity), amplification 1 (number of visits to website), amplification 2 (number of unique visitors to website), leadership (proportion of study activities offered that were attended by participants), community mobilization (numbers of participants in annual advocacy event), social media influence 1 (number of clicks on social media content), social media influence 2 (number of new content posted/month), intersectional stigma reduction 1 (proportion of staff participating in training sessions), intersectional stigma reduction 2 (proportion of eligible providers participating in training sessions), peer support (proportion of participants using peer support services). For optimization, the delivered doses of each strategy will be measured at least once before the end of each LAGO stage and average costs needed per unit of delivery of each strategy will be assessed, e.g. each peer support interaction costs $20 (for 1 hour of time for each interaction for technical staff to facilitate the interaction and track its occurrence).

Other ongoing funded studies of which we are aware in which LAGO will be used include the ImpleMEntation of a Digital-first Care deLiverY Model for Heart Failure in Uganda (MEDLY Uganda)[36] and the C4P Study, Client-Centered Care Coordination for Black Men Who Have Sex With Men[37,38].

**Discussion**

To conduct a LAGO study, inputs must be specified (Box 1). Should minimizing cost be an optimization criterion, as is usually the case, a critical input to the conduct of a LAGO study will include empirical or "guesstimated" average total costs as a function of the intensity of the implementation strategies to be optimized, along with some sense of the shape of the cost function and the admissible range of doses. Inputs may also include minimum power to be attained for the standard end-of-study test of the overall package effect, and/or the attainment of a certain outcome goal. Bounds for the doses of each component to be optimized are required. Then, investigators can Learn As they GO, repeatedly optimizing the implementation strategy bundle, or intervention package components, subject to the pre-specified criteria until the study's end. To do this, data must be collected on the dose of each component delivered at some time points during each LAGO stage and on the outcome or the surrogate outcome to be used for optimization. When LAGO is used, the chances of trial failure will be reduced, an optimal intervention package will be estimated with its uncertainty, overall and for subgroups if desired, information will be available on the relative impact of each implementation strategy on the outcome that was optimized, and a rule for future optimal intervention packages in new environments will be available, as long as data on the factors characterizing these new environments is collected and included in the LAGO modeling. See Box 3 for a concise summary on "Reading a LAGO trial: Questions Investigated, Outputs, and Interpretation". We look forward to future publications describing the outcomes of the application of LAGO to PULESA, TASKPEN, MAP-IT and HPTN 096.

Further work on developing LAGO methods to apply to a wide range of study settings is in progress. Extensions to clustered and longitudinal data are underway. Development of causal inference methods to adjust for possible time-invariant and time-varying confounding by facility- and patient-level covariates is of interest. User-friendly software is under development. The elucidation of methods for choosing, before a LAGO study begins, the number of LAGO stages, the number of clusters, the cluster sizes, and the sampling strategy for periodically measuring the implementation of intervention components within facilities is also underway. Methods for incorporating uncertainty in the cost estimates, and in the form of the cost function itself, are needed. As we gain experience with LAGO in the ongoing studies described above, and some others, further issues requiring methodologic solutions will undoubtedly emerge.

With the vast disparities in the rates of many major diseases between high versus low and middle income countries, as well as the disparities existing within the United States and other high income countries by race/ethnicity, gender, and other indicators of marginalization, it is evident that most of the world's most pressing health problems, including tuberculosis, HIV/AIDS, cardiometabolic disorders, maternal and under-5 mortality, mental health, and cervical, colorectal and lung cancer, can be largely eliminated using the well-known, effective interventions already available. Without any new scientific discoveries, implementation and prevention science promises to address the gap between what we know about the prevention of these diseases and what we do in practice. What is needed now are cost-effective implementation strategy bundles that allow these evidence-based interventions to be scaled up and out, in the form of comprehensive complex packages that target one or more diseases simultaneously, to address the major preventable health issues in contextually sustainable forms[39].

We who follow the scientific method as the primary form of knowledge generation may wish to consider that the engineering paradigm, which indeed involves tweaking, tailoring, and adapting of an underlying concept until an acceptable or even superlative product is attained, has brought us most of what is central to modern life: the car and plane; the telephone, television, radio and computer; electricity and light after dark, roofs that don't leak when it rains, central heating, much of modern agriculture and nutritional energy abundance; movies, videos, and mass-distributed music. None of these achievements were brought about by randomized experimentation. By using the rigorously developed LAGO design, we can leapfrog over the ill-fitting paradigm of the individually randomized clinical trial from which biomedicine has achieved so much, primarily in terms of late-stage treatment, and avoid further null results for trials studying the implementation of proven public health interventions.

Box 1. Inputs to a LAGO study

| | Parameter | Example, BetterBirth (illustrative example from[8,25,40] | Required or Optional? |
|---|---|---|---|
| Features of study site(s) | Maximum number of LAGO stages | 3 | Required |
| | Sample size per stage | 113; 2143; 5103. Final analysis (see Outputs) uses full data set of 7359 deliveries, and all 36 facilities | Required |
| | Number of clusters per stage | 2; 4; 30 | Required if study is clustered |
| | ICC | 0.062 | Required if study is clustered |
| Features of bundle to be optimized | Number of strategies to be optimized | 2 (duration of study launch, total number of coaching visits) | Required |
| | Minimum and maximum value of the dose of each strategy | [1, 5] for days of launch duration [1, 40] for number of coaching visits | Required |
| | Cost of each strategy and cost function | $800 per day of launch ($x_1$) and $170 per coaching visit ($x_2$). The cost function was $C_2(x) = 1700x_1 - 950x_1^2 + 220x_1^3 + 380x_2 - 24x_2^2 + 0.6x_2^3$ (Figure 2) | If optimal intervention aims to minimize the cost, both costs and cost function are required. For non-linear cost functions, costs at specific dose values are required. |
| | Guesstimate of the effect of each strategy | See Table 2 for examples of this | Optional, helps with determining if study goals are feasible pre-trial |
| Optimization goals | | | At least one must be chosen, but there can be more than one |
| | Target value of clinical implementation goal 1(, goal 2) | 15% decrease in the combined maternal and perinatal primary health outcome rate (maternal death, stillbirth or early neonatal death, or maternal severe complications) | |
| | Target value of implementation goal 1 | Greater than 80% of EBPs performed among those measured (Absolute goal); An increase of at least 25 percentage points from baseline proportion of performed EBPs (Relative goal) | |
| | Target value of implementation goal 2 | EBP adherence score across the 18 essential birth practices measured from the Safe Childbirth Checklist. In BetterBirth, 90% of this would be 16. Alternatively, a goal could be a 50% increase of adherence from baseline. | If continuous, e.g. average number of essential birth practices implemented at end of study, or e.g. change in average number of essential birth practices implemented from baseline to end of study, mean or mean change must be given |
| | Target value of patient and/or provider satisfaction goal 1(, goal 2) | Acceptability of the program by providers (80%) | If binary, target absolute or relative percent improvement from baseline; if continuous, numerical value goal at end of study goal, or of change from baseline, is needed |
| | Power goal | Standard test of overall intervention package effect has at least 80% power | Optional |
| | Subgroups for optimization and/or subsequent program planning | Small vs. large volume facilities; rural vs. urban facilities, or at specific values of the continuously measured subgroups, e.g. the first quartile | Optional |

Box 2. Outputs from a LAGO study

| Output quantity | Example from BetterBirth | Notes |
|---|---|---|
| P-value for the test of the overall intervention framework effect | Accounting for the matched pair clustered design in stage 3, and the pre-post design in stages 1-2, the test of the overall intervention effect gave a p-value <0.001 after adjusting for clustering, for the outcome of number of EBPs performed among those measured. | This p-value will not depend on the model used to estimate the effect of the individual intervention package components on the targeted outcomes. For example, the test for the difference between two binomial proportions may be used as the test statistic for the primary hypothesis, while a logistic regression model is used to estimate the impacts of the individual components on the outcomes targeted for optimization. |
| Optimal intervention package (95% confidence set) | For facilities with a mean birth volume of 175 births/month, the optimal intervention is a launch duration of 4 days with 36 coaching visits.<br><br>95% confidence set includes 7.5% of all possible combinations of the implementation strategies within the bounds. Alternative implementation strategy doses in this confidence set were (3,≥39), (4,33 − 37) and (5,27 - 34), where the first number is the number of launch days and the second the number of days of coaching visits. | The 95% confidence set includes all possible intervention component packages for which the confidence interval of the predicted outcome rate includes the outcome goal. Among all of these, the optimal is the one that best satisfies the optimization goal(s), here cost. |
| By subgroup | For the 25th percentile of birth volume (130 births/month), the optimal intervention is a launch duration of 4 days and 33 coaching visits. | Other subgroups that could have been considered in this context include the number of skilled birth attendants in the facility, distance to facility from patients' residences, both suitably grouped into meaningful categories, and facility level (Primary health center, Community health center, First referral unit) |
| Impact of each package component, (95% CI) | Coaching Visits (per 5 visits) and launch duration (days) increase the proportion of EBPs performed by 18.8 percentage points (95% CI 18.2, 19.3) and 18 percentage points (95% CI 12, 24), respectively. | Depending on the form of the optimization target(s), impact can be expressed as odds ratio, relative risk, mean difference, or difference between mean differences |
| By subgroup | Significant interactions of both components were evident in the data. For example, the odds ratios for launch duration (per day) and coaching visits (per 5 visits) for percentiles of birth volume were<br>$25^{th}$ (1.22, 1.03)<br>$50^{th}$ (1.03, 1.04)<br>$75^{th}$ (0.98, 1.05) | Optional |
| Predicted outcome (implementation, patient satisfaction) goal at the recommended final package (95% CI) | 0.80 (95% *CI* 0.78, 0.82) of EBPs performed per birth | It is possible that given the bounds of each strategy, there is no bundle that can achieve the goal |
| By subgroup | For the 25th percentile of birth volume (130 births/month), the predicted mean outcome, proportion of EBPs performed per birth at the recommended final package was<br>0.80 (95% CI 0.80, 0.84) | Optional |
| Cost of final recommended package (95% CI) | $16,250 (IQR $16,324 – 18,545) per facility, within the 95% confidence set of optimal interventions defined by the 95% confidence interval of the estimated outcome goal at the optimal intervention) | |
| By subgroup | At the 25th percentile of birth volume (130 births/month), the optimal intervention is a launch duration of 4 days and 33 coaching visits, leading to a total cost/facility of $13646 (IQR $13976, $16626 within the 95% confidence set of optimal interventions) | Optional |

Box 3: Reading a LAGO Trial—Questions, Outputs, and Interpretation

1. What Research Questions Does LAGO Answer? Unlike a standard trial that only answers the question, "Does this fixed intervention work?", a LAGO trial answers this question along with simultaneously addressing two additional areas of inquiry:

- The Effectiveness Question: Did the strategy of adapting and refining the intervention lead to significantly better outcomes compared to the control group?

- The Optimization Question: What is the most cost-effective combination of components and dosages required to achieve a specific goal (e.g., 80% adherence to anti-hypertensive medication)?

- The Component Question: Which specific components were the active drivers of success, and by how much?

2. How Do We Interpret the Results of a LAGO trial? In a LAGO trial, the intervention composition and "dose" may change between stages.

- The Global Test: Comparing the LAGO arm against the control arm tests the adaptive framework itself, a significant result indicates that the process of implementing, learning, and tweaking the intervention was superior to the standard of care.

- The Optimal Package: Using data from all stages, the model estimates the "recipe" that achieves the goal at the lowest cost. This also produces a 95% Confidence Set containing "runner-up" packages for whom their predicted outcome goal's 95% confidence limits contain the desired outcome goal. Some of these will cost more than the optimal package.

- Applications: Policymakers can use the Confidence Set to identify effective combinations that may be easier to implement in their specific local context. In addition, the recipe can be used to estimate optimal packages for new contexts characterized by variables that were measured and included in the optimization model.

Table 1 The three stages of the BetterBirth trial[5]

| Stage | Launch duration | Coaching Visits | Data Collection | Analysis |
|---|---|---|---|---|
| Stage 1: Initial Pilot | 3 days | Once every 1-2 weeks | EBP Adherence data from 2 centers | Estimate the impact of the initial strategy on EBP adherence levels. |
| Stage 2: Refinement | Reduced to 2 days based on stage 1 insights | Increased frequency to enhance impact to 2-3 visits/week | Expanded to 4 new centers | Re-assessment of the strategy's impact on EBP adherence and costs. |
| Stage 3: Optimization and Evaluation | Continuation at 2 days | Begin with twice weekly and tapering to monthly visits by 8 months | Full trial data including 15 new control and 15 new intervention centers | Final evaluation based on comprehensive data analysis of Stage 1-3 to identify the most effective and cost-efficient set of strategies to achieve the study's goals. |

Table 2. Stage-wise results on the impact of implementation strategy components on the proportion of EBPs performed. BetterBirth Study LAGO design*. Table adapted from [1].

|  | Stage 1 $n_1$=113 $\widehat{OR}$ (95% CI) | Stage 2 $n_1 + n_2$=2256 $\widehat{OR}$ (95% CI) | Stage 3 $n_1 + n_2 + n_3$=7342 $\widehat{OR}$ (95% CI) |
|---|---|---|---|
| Launch duration/day | 0.91 (0.72, 1.16) | 1.00 (0.90, 1.11) | 1.18 (1.12, 1.24) |
| Coaching visits/5 visits | 2.46 (2.20, 2.74) | 1.38 (1.37, 1.40) | 1.19 (1.182, 1.193) |
| Birth volume (monthly/100 births) | 0.03 (0.01, 0.11) | 0.85 (0.84, 0.86) | 0.82 (0.81, 0.82) |
|  |  |  |  |
| Estimated optimal implementation strategy bundle recommended for the next stage at mean birth volume (175 births/month) | $\hat{x}^{opt(2,n_1)} = (1, 16)$ | $\hat{x}^{opt(3,n_1+n_2)} = (1, 36)$ | $\hat{x}^{opt} = (4, 36)$ |
| Cost of estimated optimal implementation strategy bundle/facility | $3,363 | $11,540 | $16,250 |

* $\widehat{OR}$ are estimated odds ratios $exp(\hat{\beta})$; 95% CI-OR are 95% confidence intervals for the odds ratios; $n_k$ is the sample size of LAGO stage $k = 1,2,3$; $\hat{x}^{opt}$ is the recommended intervention at the end of the study; $\hat{x}^{opt(k,n_1+\cdots+n_{2k})}$ is the optimal recommended intervention at the start of stage $k$ based upon the combined data before stage $k$. In the recommended optimal interventions, the first component is the launch duration (days) and the second component is the number of coaching visits.

*Table 3:* **LAGO table for population blood pressure control co-primary outcome in PULESA-Uganda[30].** With an assumed baseline outcome rate of 22% in hypertension control and hypothesized increase to 34% under the HTN-PLUS intervention, the overall odds ratio for the total intervention package needs to be 1.83.

| Package component | Units | Bounds | Expected OR/Unit at pre-trial | Cost estimates, including personnel costs |
|---|---|---|---|---|
| Access to anti-hypertensive medicines | Number of 1 month supplies of antihypertensive drugs dispensed /hypertensive patients/month | [0, 3] | 1.5 = 50% increase in the odds of being in hypertension control for a 1 month supply prescribed to a hypertensive patient | $2.91 = cost of antihypertensive drugs dispensed / hypertensive patient/ month |
| Differentiated service delivery: (multi-month prescriptions) | Proportion of antihypertensive drugs dispensed as a multi-month prescription/hypertensive patient/month | [0,1] | 1.1 = 10% increase in the odds of patients in a facility-month being in hypertension control if all blood pressure (BP) medicines were given as a multi-month prescription | $7.09 = cost of antihypertensive drugs dispensed through multi-month dispensing / hypertensive patient /month |
| Differentiated service delivery: (community-based drug distribution) | Proportion of antihypertensive drugs dispensed through community-based drug distribution model/hypertensive patient/month | [0,1] | 1.25 = 25% increase in the odds of patients in a facility-month being in hypertension control if all BP medicines were dispensed through the community | $2.44 = cost of antihypertensive drugs dispensed through community-based service delivery / hypertensive patient /month |
| Access to BP machines | Number of visits at which blood pressure is measured /patients seen in past month | [0,1] | 2.1 ≈ twofold increase in the odds of being in hypertension control for each patient being measured for BP | $0.05 = cost of 1 BP measurement |
| Hypertension training | Hours of training /month/eligible provider | [0,10] | 1.05 = 5% increase in the odds of being in hypertension control for every hour of training given to a provider in patient's facility | $3.97 = cost of hour of training /eligible provider /month |
| Performance Improvement Program | Proportion of eligible providers participating in performance improvement planning after monthly audit & feedback on cascade metrics/ month | [0,1] | 1.3 = 30% increase in the odds of patients in a facility-month being in hypertension control if all eligible providers participated in monthly feedback review | $2.90 = cost of performance improvement program / eligible provider/ month |

Table 4: **LAGO table for viral suppression primary outcome in HPTN 096.**[35] With an assumed baseline rate of 57% viral suppression and a hypothesized increase to 67% viral suppression with intervention, the overall odds ratio for the total package needs to be 1.53.[1]

| Package Component | Definition/measure | Unit | Lower bound of range | Upper bound of range | Intervention condition uptake goal |
|---|---|---|---|---|---|
| Health Equity - Education--Facilitation | Proportion of participants who participate in at least one chapter activity | 1 | 0 | 1 | 0.9 |
| Health Equity - Education--Amplification | Number of visits to website | 10,000 | 0 | 10 | 5 |
| Health Equity - Education--Amplification | Number of unique IP addresses visiting to website | 1,000 | 0 | 10 | 5 |
| Health Equity - Leadership | Proportion of activities participant attended out of the total | 1 | 0 | 1 | 0.5 |
| Health Equity - Community Mobilization | Participation in annual advocacy event and science event | 1 | 0 | 2 | 1 |
| Social Media Influencers | Number of social media clicks/10,000 clicks + number of followers/1000 followers | | 0 | 20 | 5 |
| Social Media Influencers | Number of new contents posted per month | 1 | 0 | 31 | 31 |
| Intersectional Stigma Reduction 1 | Proportion of staff participating in ECHO sessions | | 0 | 1 | 0.9 |
| Intersectional Stigma Reduction 2 | Proportion of eligible providers participating in ECHO sessions | 1 | 0 | 1 | 0.5 |
| Virtual Peer Support | Proportion of participants using peer support service (at least one interaction) | 1 | 0 | 1 | 0.2 |
| | | | | | |
| Outcome -- HIV controlled | Viral suppression rate (%) | 1 | 0 | 1 | 57%-->67% viral suppression |

---

[1] Costs of each component to be determined after the study begins

Figure 1. Steps of a LAGO Study[1]

- Preliminary steps
    - Develop a theory of change to inform the selection of the initial intervention package, $x^{(1)}$
    - Determine optimization criteria, including desired mean outcome and lower and upper bounds on each of the intervention components
- Conduct Stage 1 of the $K$ stages
    - Collect data on intervention strategies as actually implemented, $A_j^{(1)}$, and outcomes, $Y_j^{(1)}$, $j = 1, \ldots, J$ facilities.
    - Determine costs of strategies in the implementation bundle
    - Determine recommended intervention for Stage 2, $\hat{x}^{(2)}$, based on the $A_j^{(1)}$ and outcomes, $Y_j^{(1)}$, $j = 1, \ldots, J$ facilities, and costs of Stage 1, subject to optimization criteria
- Conduct stages $k = 2, \ldots, K$ repeating the following steps:
    - Collect data on intervention strategies as actually implemented, $A_j^{(k)}$, and outcomes, $Y_j^{(k)}$, $j = 1, \ldots, J$
    - Refine costs
    - Determine recommended intervention for Stage $k + 1$, $\hat{x}^{(k+1)}$, based on the data collected up to stage $k$, $(A_j^{(1)}, Y_j^{(1)}, A_j^{(2)}, Y_j^{(2)}, \ldots, A_j^{(k)}, Y_j^{(k)})$,, $j = 1, \ldots, J$ facilities, subject to optimization criteria
- Final data analysis using data from all $K$ stages
    - Test the null hypothesis using outcome data from all $K$ stages, $(Y^{(1)}, Y^{(2)}, \ldots, Y^{(K)})$
    - Estimate the overall intervention effect and the effects of the individual components, using outcome and observed intervention component data from all $K$ stages, $(A_j^{(1)}, Y_j^{(1)}, A_j^{(2)}, Y_j^{(2)}, \ldots, A_j^{(K)}, Y_j^{(K)})$,, $j = 1, \ldots, J$, $k = 1, \ldots, K$
    - Estimate the final optimal intervention package, $\hat{x}^{(opt)}$, and its 95% confidence set, using outcomes and observed intervention data from all $K$ stages
    - Estimate the per facility cost of the optimal intervention package and the per facility cost of the other interventions contained in 95% confidence set

[1] Note: These steps apply to a multi-facility LAGO study with strategies implemented at the facility level, by providers and/or the system, as has been our experience in practice. The notation would be modified slightly for strategies implemented at the individual level, and for studies conducted in a single facility.

Figure 2. Total[2] implementation costs by change in program intensity. (Illustrative linear and nonlinear cost assumptions for launch duration and number of coaching visits in the *BetterBirth* program.)

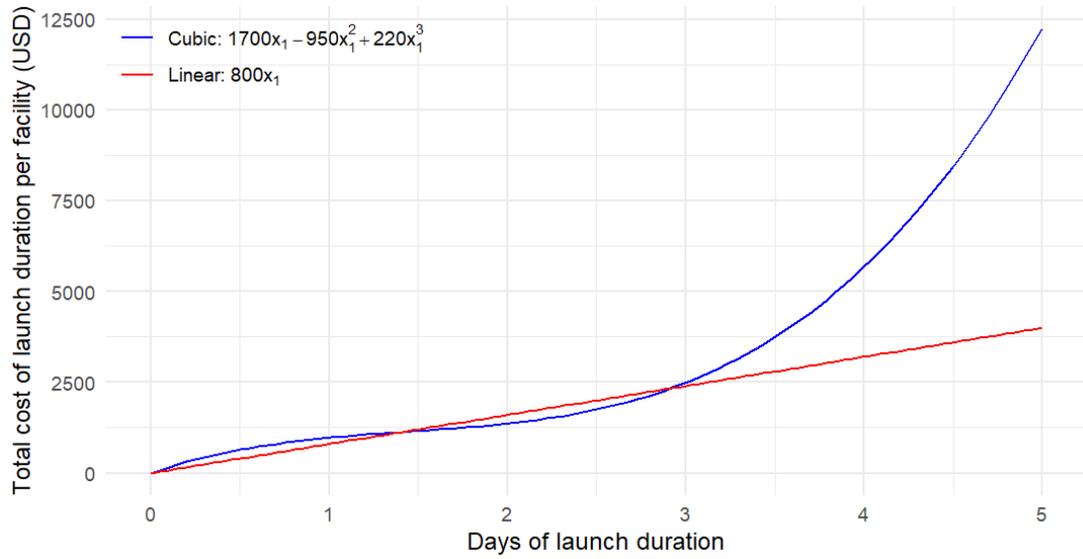

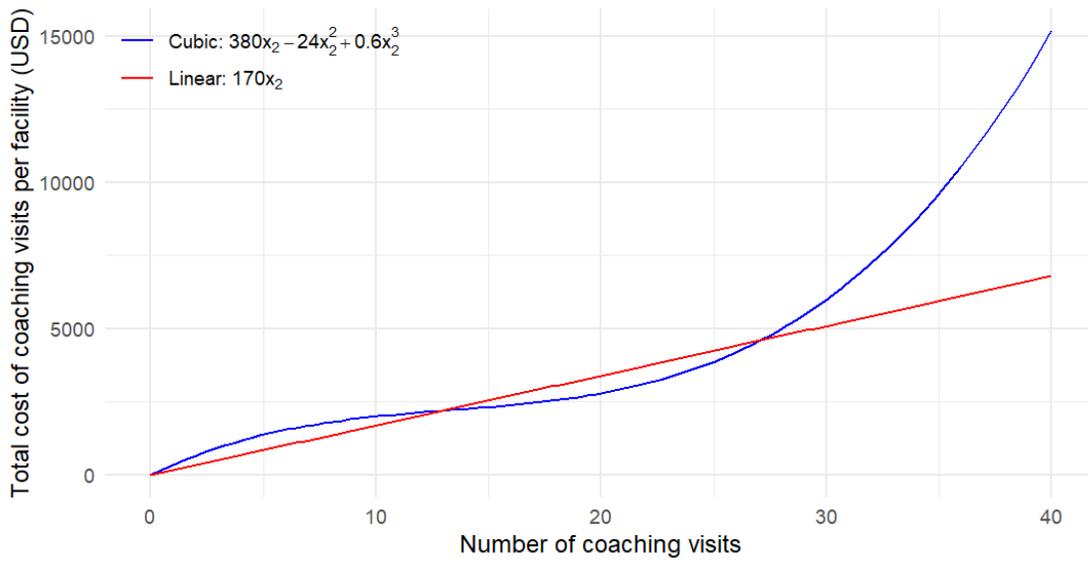

---

[2] Total cost reflects the full cost of delivering the component, including fixed start-up costs and variable costs, at a given intensity or dosage (x) in number or duration.

**Contributors and sources**

Donna Spiegelman, Sc.D., M.S., is the guarantor of this article. She drafted the manuscript and led much of the work described within. She is Susan Dwight Bliss Professor of Biostatistics and Director of the Center on Methods for Implementation and Prevention Science, Yale School of Public Health, New Haven, CT. She is the Principal Investigator of the two grants that funded this research. Judith Lok, Professor, Department of Mathematics and Statistics, Boston University, Boston, MA, USA. She is co-Principal Investigator of one of the grants that supported this work, and co-led the developments included herein. Guangyu Tong, Ph.D, Assistant Professor in the Section of Cardiovascular Medicine, Department of Internal Medicine, Yale School of Medicine, New Haven, CT, USA; Dong (Roman) Xu, Institute for Global Health and Acacia Lab for Implementation Science, Southern Medical University, Guangzhou, China; and Xin Zhou, Assistant Professor, Department of Biostatistics and Center on Methods for Implementation and Prevention Science, Yale School of Public Health, New Haven, CT, USA; each contributed to conceptualizing and revising the article. Ante Bing, B.A., now Ph.D. was a PhD. Candidate in the Boston University Department of Statistics, Boston, MA, USA. He performed the analysis included in the article, and contributed to its revisions. Jingyu Cui, PhD, Post-doctoral Fellow in the Center on Methods for Implementation and Prevention Science and Department of Biostatistics, Yale School of Public Health, New Haven, CT, US; Mona Abdo, PhD, Associate Research Scientist in the Center on Methods for Implementation and Prevention Science and Department of Biostatistics, Yale School of Public Health, New Haven, CT, US LaRon Nelson, Professor, Yale School of Nursing, New Haven, CT, USA ; Fred Semitala, MBChB, MMED, MPH-EPI, IMS, Investigator, Infectious Diseases Research Collaboration, and Lecturer, Department of Medicine, Makerere University, and Makerere University Joint AIDS Program, Kampala, Uganda;  John Baptist Kiggundu, Infectious Diseases Research Collaboration, Kampala, Uganda; Chris Longenecker, Director, Global Cardiovascular Health Program, Department of Global Health, University of Washington, Seattle, Washington, USA; and Charles Goss, Assistant Professor, Center for Biostatistics and Data Science, Washington University School of Medicine, Saint Louis, MO, USA; each contributed to the  collection and/or presentation of data included in this paper, and in the revisions.

Grants to acknowledge: R01HL167936, UH3L154501, UH3HL154501-04S1,

**Summary points**

- It is difficult if not impossible to specify the composition and constituent dosages of a complex multi-level intervention package or of an implementation strategy bundle before a study starts.
- The Learn-As-you-GO (LAGO) design allows investigators to optimize the composition and dosages of constituents as the study is being conducted at pre-specified stages, with no impact on the validity of the overall primary hypothesis test of the intervention.
- LAGO designs can prevent trial failure.
- With LAGO, cost can be minimized, and benefit maximized, where benefit can be quantified in terms of clinical and/or implementation outcomes, and/or patient satisfaction.
- Once a LAGO study has been conducted, new programs in new contexts with different facility- and patient-level features can be recommended that will be optimal for these.

Citations

Lusaka, Zambia: protocol for the TASKPEN hybrid effectiveness-implementation stepped wedge cluster randomized trial. *Implement Sci Commun*. 2024;5(1):61.